\documentstyle[12pt]{article}
\author{Hans - J\"urgen Schmidt}
\title{The Einstein equation should be divided by two
}
\date{}
\begin{document}
\maketitle

\centerline{
Universit\"at Potsdam, Institut f\"ur Mathematik,
Projektgruppe 
Kosmologie}
\centerline{
      D-14415 POTSDAM, PF 601553, Am Neuen Palais 10, Germany}

\begin{abstract} 
We present three reasons for rewriting the Einstein 
equation. The new version is physically equivalent 
but geometrically more clear. 
1.	We write $4 \pi$ instead of $8 \pi$ at the r.h.s,
and we explain how this factor enters as surface area
of the unit 2--sphere.
2. We define the Riemann curvature tensor and its
contractions 
(including the Einstein tensor at the l.h.s.) 
with one half of its usual value. This compensates
not only for the change made at the r.h.s., but
 it gives the result that the curvature scalar of the 
unit 2--sphere equals one, i.e., in two dimensions,
now the Gaussian curvature and the Ricci scalar coincide.
3. For the commutator $[u,v]$ of the vector fields 
$u$ and $v$ we prefer to write
(because of the analogy with the antisymmetrization 
of tensors)
$$ 
[u,v] \ = \ \frac{1}{2} \, ( \, u\, v \ - \ v \, u \, )
$$ 
which is one half of the usual value. Then, the curvature
operator defined by
$$ \nabla_{[u}  \  \nabla_{v]} \quad - \quad \nabla_{[u, \,v]} 
$$
(where $\nabla $ denotes the covariant derivative) is 
consistent with point 2, i.e., it equals one half of 
the usual value.   
\end{abstract}

\section{Introduction}

In 1914, the foundation of General Relativity Theory was
essentially finished [1]. It was completed by H. Lorentz and
D. Hilbert in 1915 by noting that the Einstein tensor
represents the variational derivative of the Riemannian
curvature scalar [2]. Einstein himself was happy
 of having finished his work from the principal
point of view, and he was aware, see e.g. [3], that
there are yet misleading notational points.

 I agree with the opinion formulated 
in the  introduction
of [4] that one should try ''to develop 
gravitational theory in the most logical and 
straightforward way - in the way it would have
developed without Einstein's intervention.''
However, I would formulate it 
 in a more respectful manner; and 
 that goal seems not to be reached in [4],
cf. [5].

The purpose of the present note is to
rewrite the Einstein equation in  a manner
 which is on the route Einstein went;
it will be physically equivalent, but it gives more
geometrical insight: We replace $8\pi$ at the r.h.s.
by $4\pi$ and show, using [6], 
 that this factor is just the
surface area of a unit sphere. 
 For comparison, we show, how several textbooks 
[7 - 11] deduce the factor $8\pi$.

As a byproduct of our form of the Einstein
equation, two
further inconsistencies of Riemannian geometry
 disappear.

 Einstein always used the lower index position for
coordinates; maybe, he was afraid of ambiguities when writing
$x\sp 2$. In [3], however, he already observed this to be an
inconsistency (''Of course, according to this definition
the $dx_{\nu}$ are components of a contravariant 
vector; however, here we continue to apply
a beloved usage to write a subscript.'') 
 So it is consequential that now
almost
 all textbooks on General Relativity e.g. [7 - 11]
use the upper position: then the contravariant 
vector $dx\sp k$ carries, 
as all contravariant vectors do, one index in 
upper position. As one advantage of this, the 
Einstein sum
convention can be made more rigorous: automatic 
summation is
performed with all indices appearing in one 
upper and in one
lower position. The line element $ds$ is then
 defined via
$ds\sp 2 \, = \, g_{jk} \, dx\sp j \, dx\sp k$. 

We mentioned this example to show that one can
change the representation of the Einstein 
equation without damaging the 
 ''spirit of General Relativity''. 
  
\bigskip

Having seen this development we feel free rewriting the
Einstein equation as follows.

\bigskip

\section{Einstein's equation \, -  \, right-hand side}

The right--hand side of the Einstein equation reads 
$\kappa \, T_{ij}$. In units where the light velocity 
$c\, = \, 1$, one takes usually 
$\kappa \, = \, 8\, \pi \, G$; 
$G$ being Newton's constant  
and $T_{ij}$ is the 
 energy-momentum tensor. We want to give an argument why   
$\kappa \, = \, 4\, \pi \, G$ is more natural. 

 Ludolf's number $\pi$ is defined as the surface of a unit
circle, and then $4\, \pi$ turns out to be the surface of a
unit sphere. Newton's constant $G$ is defined by
the acceleration $a=GM/r^2$ stemming from the 
gradient of the potential 
\begin{equation} 
\phi \ = \ - \ G \, M \, / \, r
\end{equation} 
where $\phi$ is the potential of a point mass $M$ at distance
$r$. This equation is equivalent to the Poisson equation
\begin{equation} 
\Delta \, \phi \ = \ 4 \, \pi \, G \, \rho 
\end{equation} 
where $\rho $ represents the matter density. 
Looking into the proof that relates eq. (1) to eq. 
(2) one can see that the factor $4\, \pi $ in eq. (2)  
is just the surface of the unit sphere. 

This can be done in different ways: First, one can 
prove that in the sense of distributions
$$
\Delta (-\frac{1}{r}) \, = \, 4 \pi \delta
$$
and the proof uses the surface integral over a small 
sphere; 
second, one can approach the point mass by a sequence 
of spherical shells of matter with the same result.
Third, one could look for higher dimensions
whether the factor $4\pi$ is only accidentally
 equal to the surface area.

Let $\omega_n$ be the surface of the unit sphere S$^{n-1}$ 
in the Euclidean R$^n$. It holds 
$\omega_n = 2 \pi^{n/2}/\Gamma(\frac{n}{2})$ .

Newton's constant $G$ in $n$ spatial dimensions 
 is defined by
the acceleration $a=GM/r^{n-1}$ stemming from the 
gradient of the potential 
\begin{equation} 
\phi \ = \ - \ \frac{1}{n-2}\, G \, M / \, r^{n-2} .
\end{equation} 
 This equation is equivalent (see [6]
 I ch. 5 and II ch. 4)
to the Poisson equation
\begin{equation} 
\Delta \, \phi \ = \ \omega_n \,  \, G \, \rho 
\end{equation} 
Eq. (4) clearly shows that $4\pi$ in eq. (2) is
not only by accident the surface of the sphere.
Again, one can 
prove this in the sense of distributions
by applying 
$$
\Delta (-\frac{1}{r^{n-2}}) \, = \, (n-2)
\  \omega_n  \ \delta .
$$
Using the symbol $n!$ as usual (i.e., $0!=1$,
 $n!= (n-1)! \cdot n$ ), we get
$$\omega_{2n} = 2\pi^n/(n-1)!
$$
and
$$
\omega_{2n+1} = 2 \pi^n \cdot 4^n \cdot n!/(2n)! .
$$

\bigskip

Einstein tried to generalize just this Poisson equation (2),
and he used $\rho \, = \, T_{00}$. So it is natural to write
the right--hand side of the Einstein equation as
\begin{equation}
4\ \pi \ G \ T_{ij}
\end{equation}

\bigskip

Let us now look how textbooks get the value $8\pi$:
(we do not mention those books which do not 
comment this choice). In [7], chapter 9.1 one reads
for the deduction of the l.h.s. of the Einstein
equation 
''There is only one tensor \dots namely
$E_{ij} = R_{ij} - \frac{1}{2} g_{ij} R$''. 
Looking into 
the details, one can see that this means 
''Up to a constant multiple, there is only one \dots''.
Putting this constant to 1, the $8\pi$ at the
r.h.s. is fixed. (The deduction
in [9] via eq. (IV,3,3)
turns out to be quite simular).
 In [10], the ansatz eq.(11.4)
$E_{ij} = R_{ij} + c_1 R g_{ij} + c_2 g_{ij}$
is called the only possible expression 
 (where, of course, a factor $c_0$ in front of 
$R_{ij}$ would be possible, too). 
In all these cases, an additional 
 factor $\frac{1}{2}$ in front of the 
l.h.s would lead to the r.h.s. (5) $ = 4 \pi G T_{ij}$.
   Such a r.h.s. is called an 
''apparently natural equation'' 
([11], after eq. (3.12), however, in a slightly different
context). 

\bigskip

How Einstein deduced his equation with the factor 
$8\pi$? [1] page 1076 reads: ''Wir setzen \dots, indem wir
\"uber die Konstante willk\"urlich verf\"ugen, \dots
$ H = - g^{ij} \Gamma^k_{il} \Gamma^l_{jk} $''
(We arbitrarily fix the constant such that \dots) 
  At that moment, it seems, he had already the 
knowledge about the sign, but not about the 
detailed consequences, and so he put the 
constant to 1. In eq. (73) he defines the object
$E_{ij}$ (now called the Einstein tensor)
 from the derivatives of $H$, and
again the factor in front of it was chosen to 
be 1 for simplicity. Then the famous equation (74)
$E_{ij}=\kappa T_{ij}$ follows. At page 1083 he deduces the 
Newtonian limit and writes 
$\frac{\kappa}{2} = 4 \pi G$.
	Surely, he felt at that moment, that an additional 
factor $\frac{1}{2}$ in the definition (73) of $E_{ij}$
 would more directly lead to the desired result 
$\kappa = 4 \pi G$. However, we do not know why
he did not insert it. 

\bigskip 

\section{The curvature tensor}

The antisymmetrization brackets $[ \ ]$ are defined by
\begin{equation} 
S_{[ij]} \ = \ \frac{1}{2} \, (\, S_{i \, j} \ - \ S_{j \, i}
\, ) .
\end{equation} 
The factor $\frac{1}{2}$ follows from the natural requirement
of idempotency of the antisymmetrization operator:
antisymmetrization should not alter antisymmetric tensors. 

The same kind of brackets $[ \ ]$ are used to express the
commutator $[u,v]$ of the vector fields $u$ and $v$. We prefer
to write
\begin{equation} 
[u,v] \ = \ \frac{1}{2} \, ( \, u\, v \ - \ v \, u \, )
\end{equation} 
which is one half of the usual value. This can be motivated as
follows: let $e_A$ be an $n$-bein, i.e., an anholonomic basis
in the $n$-dimensional (Pseudo-) Riemannian manifold, 
then eqs. (6) and (7) imply the validity of
\begin{equation}
  [e_{A}, \, e_{B}] \quad = \quad e_{[A} \, e_{B]} 
\end{equation}
which would be not only less aesthetic 
but also confusing 
if it needs an additional factor 2 at the right--hand side.
(What happens if we write the commutator of fields as usual,
e.g. in [8]? From (2.6) and  
(2.65)  one gets an unexplained factor
2 in (2.68).)

Now, we define the curvature operator by
\begin{equation}
 \nabla_{[u}  \  \nabla_{v]} \quad - \quad \nabla_{[u, \,v]}  
\end{equation} 
where $\nabla $ denotes the covariant derivative.
 The first term most naturally defines curvature 
from the commutator of covariant derivatives, the second 
term is deduced as follows: It is the only multiple of
$\nabla_{[u, \,v]}$ which realizes the 
requirement that  the 
curvature operator is linear with respect to multiplication
of $u$ or $v$ with scalar functions.

 In a coordinate basis eq. (9) reads
$$R\sp  k _{\ mij} \ = \ \Gamma\sp k _{l[i} \ \Gamma \sp l
_{j]m} 
\ - \ \Gamma \sp k _{m[i,j]} $$
This represents just one half of its 
usual value, see e.g. eqs. (2.18) and (2.20) 
 in [11] resp.

We keep the relations $R_{ij} \, = \, R\sp m _{\ imj}$ and $R
\, = \, g\sp{ij} \, R_{ij}$. These formulas do not depend on
the dimension, and they have the advantage that for $n=2$, $\
R$ is equal to the Gaussian curvature $K$. This is more 
 satisfactory 
 than the usual equation  $R \, = \, 2 \, K$. In our
convention, the unit sphere has $R \, = \, 1$.

\bigskip

\section{Einstein's equation \, -  \, left-hand side}

We keep the formula 
$$E_{ij} \ = \ R_{ij} \ - \ \frac{R}{2} \, g_{ij}$$ 
for the Einstein tensor, but we use the definitions of sct. 3,
i.e., one half of the usual values. This compensates for the
factor $\frac{1}{2}$ introduced in sct. 2, and it has the
advantage that now in the weak-field limit, 
$$E_{00} \ = \ \Delta \, \phi 
\ = \ 4 \, \pi \, G \, \rho \ = \ 4 \, \pi \, G \, T_{00} 
$$ 

\bigskip

\section{Summary}

Poisson's equation $\Delta \, \phi \ = \ 4 \, \pi \, G \, \rho
$ is generalized to the Einstein equation 
$$E_{ij} \ = \ 4\, \pi \, G \, T_{ij}$$
where the Einstein tensor has one half of its usual value. 
In the weak--field limit we have:
the relation of the right--hand sides is $\rho \ = \ T_{00}$,
and for the left--hand 
sides $\,ds \ = \ (1\, + \, \phi ) dt \,$ for purely 
temporal distances, see [1, p. 1084]. So, a lot of 
superfluous and embarrassing 
factors 2 have been cancelled. Going this way, the Einstein
tensor is unique, and not only uniquely defined up to a
constant factor. 

The differential form of 
energy--momentum conservation
$T^{ij}_{ \ ;j} =0$ can be proven to follow from the 
Bianchi identity, but a more lucid and direct proof
uses the fact that the Einstein tensor represents  the
 variational derivative of a scalar. 

\bigskip

The two byproducts promised in the introduction 
are: the consistency of eq. (8), and the definition
of the curvature scalar such that it now coincides 
with the Gaussian curvature in two dimensions.

\bigskip

We do not need to fix any further sign conventions: 

1. The metric signatures (- + + +) and (+ - - -) go
into each other by the transformation
$g_{ij} \longrightarrow \, - \, g_{ij}$, and 
$E_{ij}$ and all other essential quantities 
are invariant by this transformation. 

2. The mentioned condition that the curvature scalar 
of the 2--sphere equals 1 already fixes the sign conventions 
for $R$ and $R_{ij}$. (We need, of course, the 
additional, but always fulfilled, condition, that 
neither signature nor dimension explicitly enter the
definition of curvature.) 

3. A sign convention for the Weyl tensor is never
necessary - it enters always with its square. 

\bigskip

Let us conclude with a more general remark: 
The new version of the Einstein equation 
 deduced here is the result of 
more than one decade of 
 analysis of typical errors and typical 
barriers of understanding. 
  The fact that the old version is in common
use and should not be altered does not count:
 Nowadays even century--old 
mistranslations of the Holy Bible 
 will be corrected, and the just now
finished  reform
of the German language will hopefully
cancel a lot of illogicalities, why 
should not also
General Relativity Theory be freed from such burdens ?

{\Large 
{\bf References}}

\noindent 
[1] A. Einstein, Sitzungsberichte der k\"oniglich preussischen
Akademie der Wissenschaften, Berlin, pp. 1030 - 1085 "Die
formale Grundlage der allgemeinen Relativit\"atstheorie"
vorgelegt (= submitted) am 29. October 1914, ausgegeben (=
appeared) am 26. November 1914.

\noindent 
[2] ditto, pp. 1111 - 1116 "Hamiltonsches Prinzip und
allgemeine Relativit\"atstheorie" 1916. Here, the
contributions of Lorentz and Hilbert are acknowledged.

\noindent 
[3] Ref. [1], p. 1035 "Nat\"urlich sind gem\"a\ss \ dieser
Definition die $dx_{\nu}$ selbst Komponenten eines
kontravarianten Vierervektors; trotzdem wollen wir hier, der
Gewohnheit zuliebe, den Index unten belassen."

\noindent 
[4] Ohanian, H., Ruffini, R. (1994). {\it  
Gravitation and Spacetime} (Norton, New York).

\noindent 
[5] Krasi\'nski, A. (1995). {\it Class. Quant. Grav.}
 {\bf 12}, 2361. 

\noindent 
[6] Courant, R., Hilbert, D. (1924/1937). 
 {\it Methoden der mathematischen Physik I/II}
(Springer, Berlin).

\noindent 
[7] Stephani, H. (1982).
{\it General Relativity} (Cambridge University Press).

\noindent 
[8] Schmutzer, E. (Ed.) (1980) {\it Exact Solutions of
Einstein's Field Equations} (Verl. d. Wiss. Berlin).

\noindent 
[9] Schmutzer, E. (1968) {\it Relativistische Physik}
 (Teubner Leipzig).

\noindent 
[10] M\o ller, C. (1972) {\it The theory of relativity}
(Clarendon Oxford). 

\noindent 
[11] Hawking, S., Ellis, G. (1973) {\it The large scale 
structure of space--time} (Cambridge University Press).

\end{document}